# Evidence for Near-Equatorial Deposition by Energetic Electrons in the Ionospheric F-layer

SUVOROVA Alla[1,2,a], DMITRIEV Alexei[1,2,b] and TSAI Lung-C[1,c]

[1]National Central University, No.300, Jungda Rd., Jhongli, 32001, Taiwan

[2]Moscow State University, Leninskie Gory, Moscow, 119234, Russia

[a]suvorova_alla@yahoo.com, [b]alexei_dmitriev@yahoo.com, [c]lctsai@csrsr.ncu.edu.tw

**Abstract.** Near-equatorial great enhancements of the quasi-trapped >30 keV electrons at low-altitudes are studied during strong geomagnetic storms. We have found that the transient phenomenon often began at the morning and was capable to live several hours predominantly over the Pacific region with flux intensities largely exceeded quiet-time level by 4-5 orders of magnitudes. The purpose of the study is to validate an assumption that the enhanced quasi-trapped electron fluxes can be an important source of the ionization in the topside ionosphere in the range between the low and middle latitudes and to estimate energy deposition. We show that ionizing particle effect can significantly contribute in storm-time increase of total electron content (TEC) or positive ionospheric storm. We present analysis of medium/great flux events during moderate/major storms on 22 July 2009 and 27 July 2004 as examples for demonstrating concomitances of enhanced electron fluxes with ionospheric storm positive phases. Addition experimental support concerning the F2-region rise was obtained from COSMIC/FS3 radio occultation data for the storm of 22 July 2009.

**Introduction**

Coupling between the magnetosphere and the ionosphere through precipitating particles from radiation belts and central plasma sheet, which strongly controls the ionization and conductance in the ionosphere, is known for a long time [1]. The energy deposition of precipitating particles with different energies mainly through the ionization of the atmospheric atoms occurs at different altitudes, particularly in the D, E, and F region of the ionosphere and extends from high to middle latitudes and over South Atlantic Anomaly (SAA) at low latitudes [2-7]. Some signatures of the particle effect were found at middle latitudes, particularly uplift of the F region ionosphere, spots of enhanced ionization at the F-layer bottom heights, and also in the topside F region over SAA. The contribution of the particle effect in the total electron content (TEC) increase at middle latitudes is about 10 TECU (1TECU=$10^{12}$ el/cm$^2$) [4].

Information about storm-time particle energy spectra, from which the energy flux is derived, is now well-known for the populations in the radiation belt and auroral zone, i.e. mainly at L shells >2. It was established that during geomagnetically active periods, 1-30 keV electron fluxes in the auroral/sub-auroral regions increase to $10^5$-$10^7$ el/(cm$^2$ s sr), hereinafter el/(cm$^2$ s sr) will be referred to as units, and energy flux increases from 1 to 20 mW/m$^2$.

It is widely accepted that main sources of the energetic electrons in the inner magnetosphere at ionospheric altitudes from ~100 to ~2000 km are the inner and the outer radiation belts (IRB and ORB, respectively). The bottom of the IRB at the equator locates at an L-shell of ~1.2 and its height changes from 1600 to ~ 300 km in dependence on the longitude. There are three populations of particles, trapped, precipitating, and quasi-trapped, classified on physical behavior of particles with different local equatorial pitch angles (an angle between the velocity of a particle and the magnetic field line) [8]. In particular, particles from "quasi-trapped" population cannot close the full drift around the Earth and pitch angles of the quasi-trapped particles range within a drift loss cone. These particles can make a number of bounces but at a certain longitude, their local equatorial pitch angle occurs within a bounce loss cone and finally the particles precipitate in the upper atmosphere.

The quasi-trapped population of energetic electrons is observed below the IRB at L < 1.2 in the near-equatorial region. While its existence was known long time ago [9-11], the information about the fluxes and spectra was scarce and controversial. Traditionally it is thought that the electron fluxes below the IRB are invariably weak and certainly less than inside the IRB zone, therefore the particle impact is insufficient to produce appreciable ionization [1]. On the other hand, from the second spaceship experiment Savenko et al. [12] found sporadic events with unusually large fluxes of ~10 keV quasi-trapped electrons in the equatorial ionospheric F-region at ~320 km in the region between 150°E and 150°W longitudes. Later, Leiu et al. [13] found an appearance of the energy spectrum peak at ~10 keV of the precipitating low-energy electrons during a disturbed period. The flux increased about ~100 times greater than the quiet level and 10 times greater than the flux in the SAA. These features were observed at ~ 240 km height over Indochina and Pacific regions and located below the edge of the IRB at L~1.16, i.e. coincide with the areas revealed by [12].

Modern experiments measured the energetic electrons fluxes below and inside the IRB [14-17]. Evans D.S. [18] first have reported about dramatic enhancements of the energetic electrons fluxes at ~900 km heights in response to large magnetic storms, thereafter other studies also have noticed intense fluxes at low-altitude near equator [e.g.,19]. The observed phenomenon argue in favour to a confidence of the past observations in lower energy range and at the lower altitudes mentioned above. Note, that no explanation for the appearance of quasi-trapped electrons at so low L-values (<1.2) is yet known. Observations showed that the increases of few tens of keV quasi-trapped electrons were much larger than of precipitating electrons and comparable with the auroral precipitation intensities of $>10^6$ units, and so it is reasonably to expect a significant ionization impact in the ionosphere. However, these phenomena are not completely investigated due to lack of the information on energy spectra, pitch-angle distributions and etc.

Recently, case study and statistical analysis of the equatorial flux phenomenon at topside ionosphere heights during major geomagnetic storms has been made by Suvorova et al. [20-22]. The energy flux of electrons over the Pacific region was estimated to be 5 mW/m$^2$ during storm of 15 December 2006 and 20 mW/m$^2$ during storm of 9 November 2004. It was shown that these values allow creating an abundant ionization of a few tens TECU at the low and middle latitudes that is in a good agreement with observation of the ionospheric TEC enhancements. In the current paper we analyze case events during major storm on 26-27 July 2004 and moderate storm on 22 July 2009.

**Case events**

**Data sources.** We use time profiles of >30 keV, >100 keV and >300 keV electrons fluxes measured by polar orbiting NOAA/POES satellite fleet [8] at altitudes of ~ 850 km. In the paper we present data from detector, which measures quasi-trapped flux at low latitudes. Experimental data about low-energy electrons (30 eV-30 keV) from polar orbiting DMSP fleet have also been used to substantiate the POES observations. The altitude of DMSP satellites was 840 km. Global ionospheric maps (GIM) were acquired from a world-wide network of ground based GPS receivers through website [23].

**Great flux enhancement on 27 July 2004.** A CME-driven storm on 26 - 27 July 2004 was started ~23 UT on 26 July. A local peak of storm activity was detected at ~02 UT and the main peak at ~12 UT on 27 July [24]. Northward turning of Bz IMF after 2 UT was followed by sharp rise of the solar wind dynamic pressure to about 30-40 nPa, which caused strong magnetospheric compression with decreasing of dayside size to around ~7 Earth' radius as estimated from a model [25].

During the main and partial recovery phase 0-6 UT accompanied by the magnetospheric compression, POES satellites detected great enhancements of the quasi-trapped electrons in two energy ranges >30 keV and >100 keV near equatorial latitudes, ±30° [22]. At this time, the geomagnetic activity significantly weakened due to the abrupt northward turning IMF. Fig. 1 (left) shows great enhancements of the mid-energy electrons (>30 keV) in very wide longitudinal range extended from 30°E east to the SAA region and all local times are clearly seen against the low background intensity observed during other times in presented 36-hours period. We believe that appearance of particles at low altitudes caused by the strong compression of the magnetosphere.

The mid-energy electron fluxes have peaks around $10^7$ units in both hemispheres, while the high-energy electrons (>100 keV) have maximal intensity of ~$10^7$ units only over the night-time SAA area. The enhancements of the >30 keV electrons in the Eastern hemisphere encircled in the Fig. 1 (left) were observed during ~15 min passes of the low-latitude region at ~0040 UT (lon. 90°; 7LT), ~0220 UT (lon. 60°; 7LT), ~0250 UT (lon. 120°; 10LT), ~0345 UT (lon. 160°; 14LT), and ~0530 UT (lon. 130°; 14LT). Smooth shape of the time profiles and limited lifetime of the near-equatorial enhancements infer non-sporadic electron penetration to the topside ionosphere and gradual and relatively fast radial transport of the electrons in the magnetosphere [22].

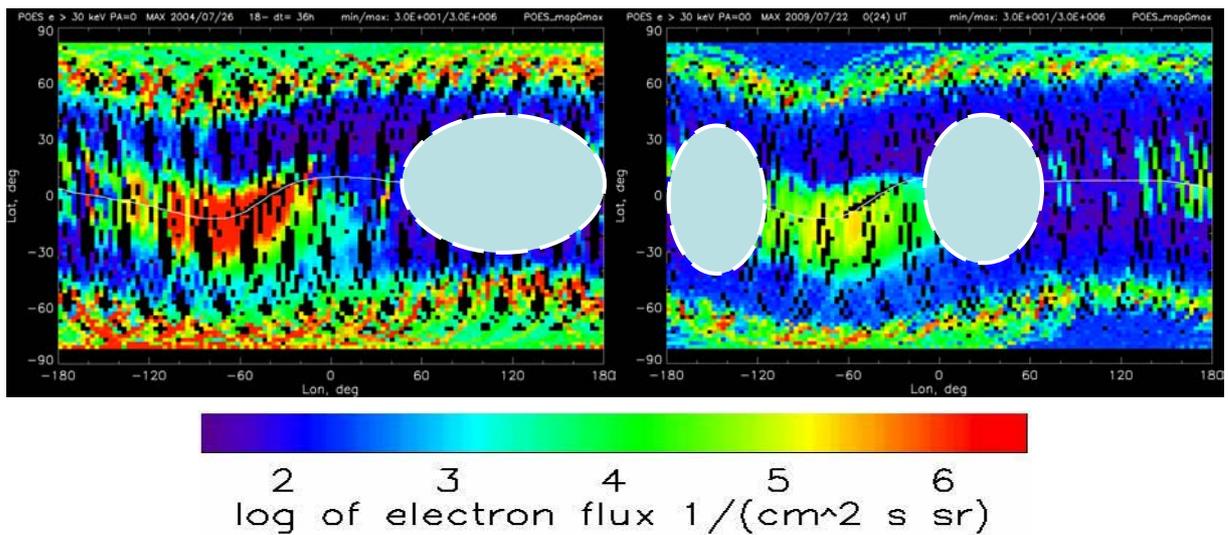

**Figure 1.** Geographic map of storm-time >30 keV quasi-trapped electron fluxes: (left) on 26-27 July 2004 and (right) on 22 July 2009. Electron enhancements observed in morning or pre-midnight local times are encircled (see text for details). The solid white curve indicates the geomagnetic equator.

We have fitted the integral energy spectra at different longitudes and times using the POES data about >30->300 keV electron fluxes. All spectra are descending. According to the DMSP data about 1-30 keV electrons the spectra are ascending [22]. Thus, energy peak is at about 30 keV. Electron energy flux density JE and integral energy flux JEmax were estimated using an assumption about anisotropic pith-angle distribution (see Discussion section). Table 1 presents the estimations of JE and JEmax for two time intervals and in three longitude regions (eastern, western longitudes and SAA area). For example, in time interval 3-4 UT, the enhanced fluxes in the energy range 30-100 keV produced energy fluxes of ~3.1, 0.2 and 7.1 mW/m² at the longitudes 156°E, 132°W and 78°W in the Pacific and SAA regions, respectively; at 4-5 UT and 159°W the energy flux was 3.2 mW/m². As one can see *the integral energy fluxes near equator can be comparable with auroral precipitations*. If we take decay rate of the ionospheric F region density as $\sim 5 \cdot 10^{-3}$ s$^{-1}$ (see Discussion section), we roughly estimate the electron impact to the ionization as ~29 TECU for 30-100 keV electrons and ~66 TECU for 1-30 keV electrons, totally ~95 TECU.

**Table 2.** Estimates of energy flux density JE, integral energy flux JEmax in range 30-100 keV and TEC during storm-time interval on 27 July 2004.

| Time interval | Parameter | Eastern hemisphere | | | Western hemisphere | |
|---|---|---|---|---|---|---|
| | | | | | Pacific | SAA |
| | | (POES) | (DMSP) | (Total) | (POES) | (POES) |
| 03-04 UT | Longitude [°] | 156E | 68E | -- | 132W | 78W |
| | JEmax [mW/m²] | 3.1 | 7.1 | 10.2 | 0.2 | 7.1 |
| | JE [eV/cm² s sr] | $3.1 \cdot 10^{11}$ | $7.1 \cdot 10^{11}$ | $1 \cdot 10^{12}$ | $1.8 \cdot 10^{10}$ | $7.1 \cdot 10^{11}$ |
| | TEC [TECU] | 29 | ~66 | ~95 | ~1.7 | ~65 |
| 04-05 UT | Longitude [°] | 142E | | | 159W | 30W |
| | JEmax [mW/m²] | 0.1 | - | 0.1 | 3.2 | 11.6 |
| | JE [eV/cm² s sr] | $1 \cdot 10^{10}$ | | $1 \cdot 10^{10}$ | $3.2 \cdot 10^{11}$ | $1.2 \cdot 10^{12}$ |
| | TEC [TECU] | ~1 | | ~1 | 29 | ~53 |

**Positive ionospheric storm on 27 July 2004.** Fig. 2 shows two-hour global ionospheric maps (GIM) of residual VTEC (dVTEC) from 22 UT to 04 UT. The residual dVTEC is calculated as a difference between the storm and quiet days, 26-27 July and 4 August, respectively. A method of selection for quiet-time condition is described in [20-22] (see Discussion). There is small excess ionization of ~15 TECU in the afternoon at the low latitudes mapped at 22-24 UT before the storm onset and during initial phase as seen from SYM-H index (see red line in the top panel of Fig.2). Then dVTEC is gradually amplified during the main phase development from 23 UT to 02 UT. The ionization in the crest of equatorial ionization anomaly (EIA) in the afternoon over the East Pacific region increased due to the prompt penetrating electric field (PPEF) as expected from the full electrodynamical scenario [26]. During the partial recovery of SYM-H (green line), the positive storm phase (2-4 UT and 4-6 UT maps) unexpectedly was maximized particularly from noon to dusk over the Pacific and the SAA regions (from ~100°E to 30°W). Also note, that the dVTEC increase significantly extends to the mid-latitudes and to morning and pre-midnight local times (encircled areas). *Thus, the positive storm was intensified from morning to midnight under Bz>0, despite of the mechanism of the prompt penetration electric field (PPEF) does not work at this time*. Moreover, the morning-to-midnight VTEC increase contradicts to the disturbance dynamo electric field (DDEF) mechanism during the recovery phase [27].

It is important to emphasize that *at the same regions and times, the enhanced energetic electron fluxes were observed* (compare Fig.1 and Fig.2). Note, that the spatial distribution of VTEC enhancements was not uniform and consisted of a number of spots of enhanced ionization. A separated spot around noon in the Southern hemisphere with the maximal dVTEC of ~35 TECU and maximal >30 keV electron flux of $10^7$ units were observed in the same longitudinal (150-160E) and time (2-4 UT) intervals. Narrow strips of small increases (<10 TECU) are seen at the northern and southern 20°-30° latitudes from 60°E to 100°E in the early morning hours (6-9 LT), where also enhanced particle fluxes of $10^5$-$10^6$ units were detected. Between 2 and 6 UT concomitance of the enhanced electron fluxes ~$10^5$-$10^7$ units and spots of the ionization strong increases of ~40 TECU were also revealed at the SAA longitudes and in the East Pacific sector (60-120W) at the afternoon and after the sunset. Sufficiently large dVTEC value of ~10-20 TECU and peak of electron flux ~$10^7$ units were observed over SAA at pre-midnight. We have to point out that *all these features cannot be predicted by the full electrodynamical scenario*. Note also, that the EIA crest at local noon hours was restricted to within ±15° geomagnetic latitudes and did not extend to the low or middle latitudes, as expected from the penetration electric field and/or from the equatorward neutral wind mechanisms.

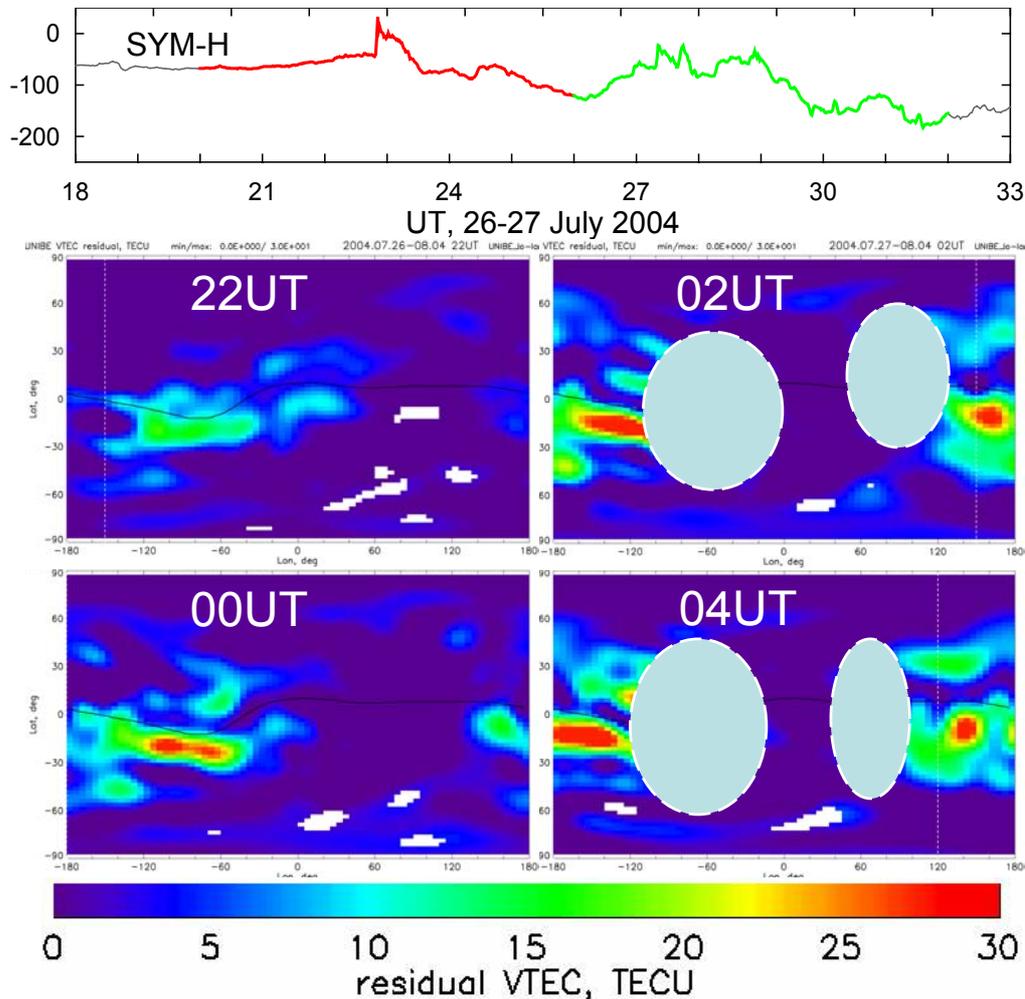

**Figure 2.** Ionospheric storm on 26-27 July 2004. Positive phase calculated as residual VTEC between storm day and quiet day (4 August). Encircled areas correspond to pre-midnight and morning local times. The solid black curve indicates geomagnetic equator, dashed vertical line indicates local noon.

Hence, *we can establish two observable facts, which clearly contradict to the electrodynamical scenario: prenoon-noon and dusk-premidnight positive storm*, when operation of the general drivers is tenuous/absent or opposite. It is quite reasonable to assume that the effect of neutral wind circulation change due to Joule heating could not play important role in the low-latitude positive storm, particularly at night in the absence of the ionizing source. Also, the effect of equatorward neutral winds should be ruled out at daytime, taking into account the sharp change of the solar wind condition in the beginning of the storm. Nevertheless, positive phase enhancement lasting for 4 hours was occurred after the magnetic storm intensification.

We consider the great enhancement of low and middle energy electron fluxes as the most likely source of the abundant ionization. We find the spatial pattern of >30 keV electron enhancements is very similar to the spatial distribution of the positive ionospheric storm (see Fig.1 and Fig.2). The time interval of the electron enhancements overlaps with the increases of VTEC. The estimated value of electron impact to the ionization ~30-100 TECU is large enough to supply and maintain the observable ionization increase by ~30-40 TECU. Hence, *we suggest that the energetic electrons can significantly contribute to the redundant ionization of the ionosphere and might be considered as an important supplement to the other general drivers of the ionosphere, especially in the early morning and night sectors.*

**Moderate flux enhancement on 22 July 2009.** Here we will address the following question: is the moderate intensity flux of quasi-trapped energetic electrons important in terms of their effects on topside ionospheric. We present moderate ionospheric density disturbances that occurred on 22 July 2009 and associated with a intense geomagnetic storm, which was classified as a CIR-driven storm [28]. Intense storm on 22 July 2009 was started near 2 UT on 22 July. The storm had a double dip pattern of the main phase with two minima of almost -95 nT at 6 and 9 UT.

Fig.1 (right) shows the >30 keV electron flux intensity distribution in the geographical coordinates along the POES satellite orbits for interval 0-24 UT. A sudden increase of three orders of magnitude in the quasi-trapped electron flux initially occurred over Africa (0-60E) at the end of the main phase (4-7 UT) in the morning sector (encircled area in the eastern hemisphere). Later, at 8-15 UT, the enhanced fluxes appear over the Asian-Pacific region at night during the early recovery phase (the encircled area in the western hemisphere over Pacific Ocean).

**Positive ionospheric storm on 22 July 2009.** Fig.3 (left panels) shows two time sequences of the GIMs, 6-8 UT and 8-10 UT, which demonstrate maximal positive storm phases, which were observed during partial recovery and second dip of SYM-H index (see red line in right-top panel of Fig.3). A day on 18 July is used as a geomagnetically quiet for dVTEC calculation. Note, that positive storm gradually develops from 0 to 10 UT and quickly decays from 10 to 16 UT. Its peak intensity hardly reaches 20 TECU at 8-10 UT. Maximal ionization is observed in the afternoon over Asia-Australia region. There is a typical crest-structure EIA at the dayside. The disturbance of the EIA perhaps associates with the PPEF or DDEF mechanisms during the main phase of the geomagnetic storm.

Electron content variations perhaps reveal some relation with the energetic electrons behaviour. Enhanced ionization encircled in the map 6-8 UT in the morning is well overlapped with the appearance of the energetic electron fluxes observed by POES satellites (see Fig. 1, right and Fig. 3, left). Those quasi-trapped particles which observed ~15° latitudes off the dip equator have mirror points at lower heights than the satellite orbit, so particles will deep penetrate into the topside ionosphere and F2 region due to the bounce motion along field line at low latitudes around 20-30° and cause the additional ionization there. It is important to note that the charged particle spend most of the time in vicinity of mirror point. Noticeable continuous enhanced ionization (~7 TECU) is observed in the night sector of longitudes 140°E-120°W from 8 to 16 UT (encircled area in 8-10 UT map). The enhancement can be related with the energetic electrons drifting a long time through this longitudinal region. Because it is occur near the midnight during the recovery phase, other sources of the ionization couldn't operate [26-27].

Additional support for particle effect has been obtained from vertical profiles of electron concentration (EC) measured in COSMIC/FS3 space-borne experiment. Six satellites of the COSMIC/FS3 mission provide a sounding of the ionosphere on the base of radio occultation (RO) technique. 3-D tomography of the EC is produced around the whole globe with a time step of 2 h and spatial grid of 5° in longitude, 1° in latitude and 5 km in height [29]. Fig. 3

(right) shows pairs of meridional cuts of EC for quiet/storm day obtained from COSMIC/FS3 3-D ionospheric tomography at 6-8 UT at 30°E longitude (local morning) and 8-10 UT at 140°W longitude (local night). It is clearly seen rising of the F layer by 100 km in comparison with quiet condition. Prominent ionization enhancement expands to higher altitudes in the topside ionosphere (up to 500 km) at morning and pre-midnight. One can see storm-time enhancement of EC in range of latitude ±40° peaked at 300 km. Finally, significant EC increase at night is seen at middle latitudes of 30–40° in northern hemisphere at 140°W longitude. *At middle latitudes and at night/morning, the presence of elevated and widely expanded EC enhancement indicate to the operation of a magnetospheric mechanism of charged particle contribution to redundant ionization of the mid-latitude ionosphere.*

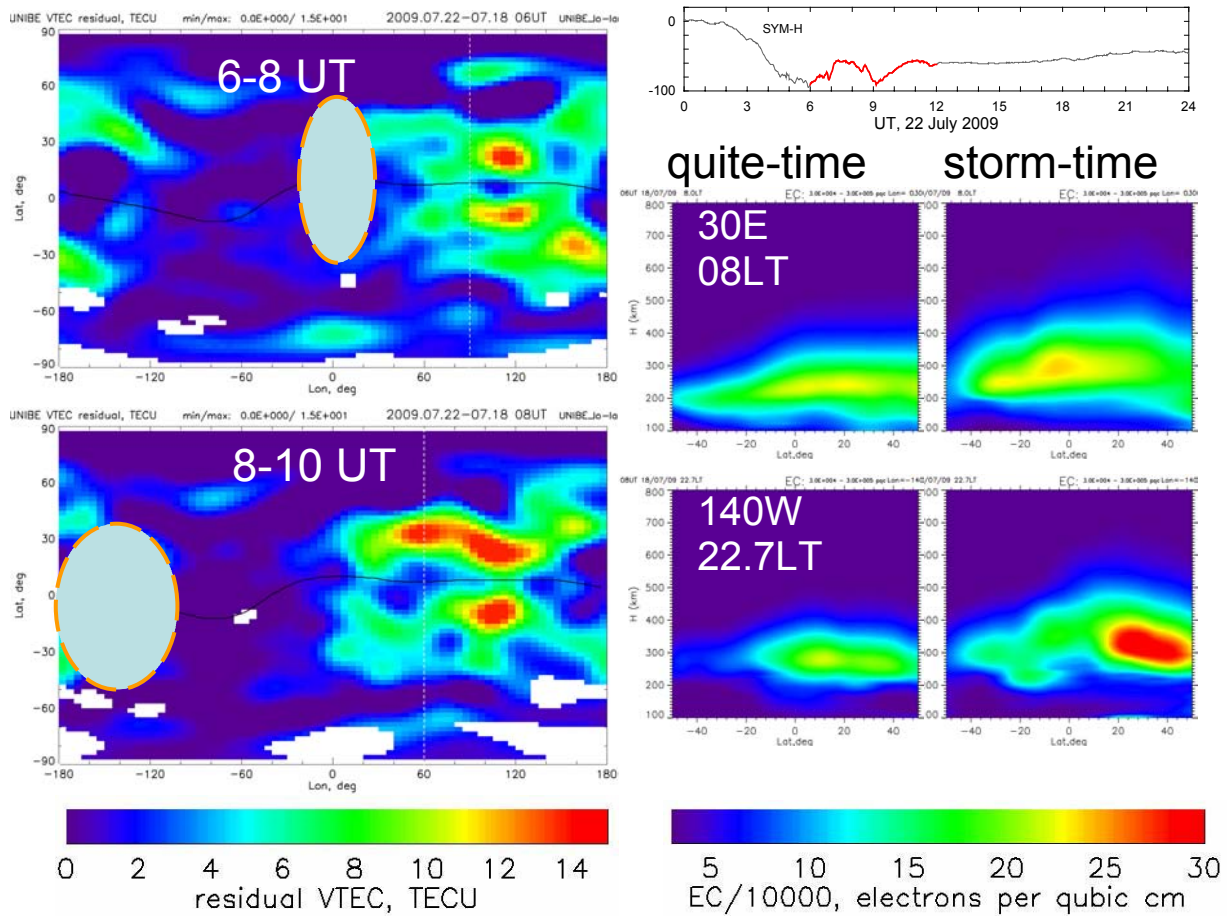

**Figure 3.** Ionospheric storm on 22 July 2009: (left) positive phase calculated as residual VTEC between storm and quiet days, encircled areas correspond to morning and pre-midnight local times; (right, bottom) COSMIC/FS3 3-D tomography of electron concentration (EC), meridional cuts of EC at longitude ~30°E at 6-8 UT and at longitude ~140°W at 8-10 UT; (right, top) the storm-time SYM-H index. 18 July was chosen as a quiet day. Geomagnetic equator is indicated by black curve. Local noon is depicted by vertical white dashed line.

**Discussion**

**Uncertainties in TEC enhancements.** We consider some uncertainties, which in total can influence by factor of two or three times to decrease or increase the resultant TEC. There are unknown actual distributions of electron pitch-angle and energy losses in such processes as the ionization, excitation, secondary electron production. In the topside ionosphere, the variability range of electron recombination rate is still not very well established. The averaged plasmaspheric TEC value and its variation is also purely known.

**Pitch-Angle Distribution.** From our study of the energetic electron flux distribution in the near-equatorial region during major storms [22], we know that particles locally trapped at a given longitude (with pitch angle ≤90°) exhibit large increases, while particles with pitch angle within the local bounce loss cone, it can be said, do not. So, we assume an anisotropic pitch-angle distribution of the electrons, and arbitrarily will use multiplying factor $2\pi$, instead $4\pi$ valid for isotropic distribution.

**Energy Losses in Ionization.** Besides of ionization, a part of primary energy of the energetic electrons is lost for excitation of the thermospheric species and for secondary electron production, but these complex processes remain to be incompletely studied. Hence, we arbitrarily assume that the electrons lose 100% of their energy for ionization of the oxygen atoms, which dominate in the topside ionosphere. Taking oxygen atom first ionization potential as 13.6 eV, we get the upper limit number of the ion-electron pairs produced by the energetic electrons. Also note efficiency of the ionization reduces with height due to thermospheric density decrease, while this effect should be offset with surplus of the long time spent by quasi-trapped particle at these altitudes.

**Impact of the Plasmasphere.** The variability of TEC measured by ground-based GPS stations relates also to plasmasphere-ionosphere coupling It is well known that the plasmasphere plays a significant role in the maintenance of the nighttime ionosphere due to the downward plasma flux, while on the dayside, the plasmasphere is filled by the upward plasma fluxes from the ionosphere, and, hence, plasmasphere certainly contributes to the ground-based GPS TEC measurements. The plasmaspheric contribution to the total electron content is typically a few TECU. The maximum effect about 5-6 TECU is observed at 2-4 LT and 14-16 LT in the low-latitudinal region, while at other local times and higher latitudes the value does not exceed 2 TECU [30]. Since the plasmasphere is depleted during a storm, its contribution also decreases. Therefore, if the expected particle-induced TEC enhancement at given local time during major storms exceeds the plasmaspheric TEC value at least by two-three times then the contribution of particle ionization is meaningful.

**Recombination Rate.** Ionization balance in the ionosphere is largely controlled by production (e.g., solar EUV flux), transport (e.g. thermospheric winds) and loss (e.g., recombination) processes. For the current study, we need to know a decay rate of the ionospheric electron density predicted by the atmospheric ion-molecule chemistry, in particular, for the height interval from 300 to 600 km in the topside F region. Basic recombination and excitation rates strongly depend upon the thermospheric species, temperature and electron density, which highly vary during geomagnetic disturbance. This fact leads to recombination rate uncertainty about a factor of two. Resent studies of the high-latitude trough [e.g., 31] showed that during intensive electron precipitation, when chemical composition and thermospheric species temperatures strongly change, the recombination rate at 300 km height can significantly increase at least 10 times compared to usual value of $2 \cdot 10^{-4}$ $s^{-1}$. Based on this and other facts, we suggest that the recombination coefficient $5 \cdot 10^{-3}$ $s^{-1}$ is a quite reasonable value and use it in our numerical estimate of TEC.

**A Problem of Quiet Days.** The geomagnetic activity practically never ceases that results in a day-to-day variability of the TEC. Generally, averaged pattern for five quietest days in a month or even a previous day before the storm onset are often used in most of ionospheric studies to calculate the storm-related TEC changes. In order to minimize the uncertainty, we use criteria for "the quietest ionosphere-magnetosphere system" based on appropriate solar wind parameters [32]. Selection method for a proper quiet day allows revealing of ionization effect (if any) of the energetic electrons with accuracy better than ~5 TECU.

**Summary**

From the case-event analysis of the magnetic storms on 26-27 July 2004 and 22 July 2009, we have demonstrated that the positive phase observed in the Pacific region at morning and pre-midnight hours during recovery phase can be explained rather by the direct ionization produced by intense fluxes of quasi-trapped energetic electrons in the topside ionosphere than by the effects of PPEF , DDEF and/or equatorward neutral winds. *We show that the energetic electron enhancements are an important source of the ionization in the topside ionosphere and, thus, they can be considered as a supplement to the general ionospheric drivers.*

The phenomenon of the enhanced quasi-trapped energetic electron fluxes is of great interest to researchers of the ionosphere-magnetosphere coupling at the low and middle latitudes. Thus, new results from the study of the enhanced quasi-trapped energetic electron fluxes and the coupling of the particle phenomenon to the TEC increases and the F region uplifts allow viewing a problem of long-duration positive ionospheric storm from a new perspective.

This work was supported by grants NSC 100-2811-M-008-093 and NSC 100-2119-M-008 -019.

**References**

[1] G.A. Paulikas, Precipitation of particles at low and middle latitudes, Rev. Geophys. Space Phys., 13(5), 709-734, (1975).

[2] J.C. Foster, M.J. Buonsanto, J.M Holt, et al., Russian-American Tomography Experiment, Int. J. Imag.Sys.Tech., 5, 148-159, (1994).

[3] M. Nishino, K. Makita, K. Yumoto, et al., Unusual ionospheric absorption characterizing energetic electron precipitation into the South Atlantic Magnetic Anomaly, Earth Planets Space, 54, 907-916, (2002).

[4] A.V. Dmitriev and H.-C. Yeh, Storm-time ionization enhancements at the topside low-latitude ionosphere, Ann. Geophys., 26, 867-876, (2008).

[5] A.V. Dmitriev, Jayachandran P.T. and Tsai L.-C., Elliptical model of cutoff boundaries for the solar energetic particles measured by POES satellites in December 2006, J. Geophys. Res. 115, A12244, (2010).

[6] A.V. Dmitriev, H.-C. Yeh, M.I. Panasyuk, et al., Latitudinal profile of UV nightglow and electron precipitations, Planetary and Space Science, 59, (2011).

[7] V.E. Kunitsyn, E.S. Andreeva, L.-C. Tsai, et al., Ionospheric imaging using various radio tomographic systems, in Proc. of XXIXth URSI General Assembly, Chicago, USA, 7-16 Aug. (2008).

[8] C.J. Rodger, M.A. Clilverd, J.C. Green, and M.M. Lam, Use of POES SEM 2 observations to examine radiation belt dynamics and energetic electron precipitation into the atmosphere, J. Geophys. Res., 115, A04202, (2010).

[9] V.I. Krasovskii, Kushner Yu. M., Bordovskii G. A., Zakharov G. F., and Svetlitskii E. M., The observation of corpuscles by means of the third artificial earth satellite (in Russian), Iskusstvennye Sputniki Zemli, 2, 59-60, (1958). (English translation: Planet. Space Sci., 5, 248-249, 1961).

[10] W.J. Heikkila, Soft particle fluxes near the equator, J. Geophys. Res., 76, 1076-1078, (1971).

[11] S. Hayakawa, Kato F., Khono T., Murakami T., et al., Existence of geomagnetically trapped electrons at altitudes below the inner radiation belt, J. Geophys. Res., 78, 2341-2343, (1973).

[12] I.A. Savenko, Shavrin P.I. and Pisarenko N.F., Soft particle radiation at an altitude of 320 km in the latitudes near the equator (in Russian), Iskusstvennye Sputniki Zemli, N13, 75-80, (1962). (English translation: Planet. Space Sci., 11, 431-436, 1963).

[13] R. Lieu, Watermann J., Wilhelm K., Quenby J.J. and Axford W.I., Observations of low-latitude electron precipitation, J. Geophys. Res., 93, 4131-4133, (1988).

[14] K. Kudela, Matisin J., Shuiskaya F.K., et al., Inner zone electron peaks observed by the ''Active" satellite, J. Geophys. Res., 97, 8681-8683, (1992).

[15] E.A. Grachev, Grigoryan O.R., Klimov S.I., et al., Altitude distribution analysis of electron fluxes at L~1.2–1.8, Adv. Space Res, 36, 1992-1996, (2005).

[16] O. Grigoryan, Kudela K., Rothkaehl H. and Sheveleva V., The electron formations under the radiation belts at L shell 1.2-1.9, Adv. Space Res., 41, 81-85, (2008).

[17] J.A. Sauvaud, Moreau T., Maggiolo R., et al., High-energy electron detection onboard DEMETER: The IDP spectrometer description and first results on the inner belt, Planet. Space Sci., 54, 512, (2006).

[18] D.S. Evans, Dramatic increases in the flux of >30 keV electrons at very low L-values in the onset of large geomagnetic storms, EOS Trans., 69(44), 1393, (1988).

[19] Y. Tanaka, Nishino M. and Iwata A., Magnetic storm-related energetic electrons and magnetospheric electric fields penetrating into the low-latitude magnetosphere (L~1.5), Planet. Space Sci., 38(8), 1051-1059, (1990).

[20] A.V. Suvorova, Tsai L.-C. and Dmitriev A.V., On relation between mid-latitude ionospheric ionization and quasi-trapped energetic electrons during 15 December 2006 magnetic storm, Planet. Space Sci., 60, 363-369, (2012).

[21] A.V. Suvorova, Tsai L.-C. and Dmitriev A.V., TEC enhancement due to energetic electrons above Taiwan and the West Pacific. TAO (in press).

[22] A.V. Suvorova, Tsai L.-C. and Dmitriev A.V., On magnetospheric source for positive ionospheric storms. Sun and Geosphere, 7(2), 91-96, (2012), http://www.sungeosphere.org/

[23] Information on ftp://ftp.unibe.ch/aiub/CODE/

[24] J. Zhang, I.G. Richardson, D.F. Webb, et al., Solar and interplanetary sources of major geomagnetic storms (Dst<100 nT) during 1996–2005, J. Geophys. Res., 112, A10102, (2007).

[25] A.V. Suvorova, Dmitriev A.V., Kuznetsov S.N., Dayside magnetopause models, Radiation Measurements, 30, 5, 687-692, (1999).

[26] B.G. Fejer, Low latitude storm time ionospheric electrodynamics (2002), J. Atmos. Sol. Terr. Phys., 64, 1401–1408, (2002).

[27] C.-M. Huang, A.D. Richmond and M.-Q. Chen, Theoretical effects of geomagnetic activity on low-latitude ionospheric electric fields, J. Geophys. Res., 110, A05312, (2005).

[28] M.-C. Fok, N. Buzulukova, S. H. Chen, et al., Simulation and TWINS observations of the 22 July 2009 storm, J. Geophys. Res., 115, A12231, (2010).

[29] L.-C. Tsai, Tsai W.H., Chou J.Y., Liu C.H., Ionospheric tomography of the reference GPS/MET experiment through the IRI model, Terrestrial Atmospheric and Oceanic Sciences 17, 263–276, (2006).

[30] E. Yizengaw, M.B. Moldwin, D. Galvan, et al., The global plasmaspheric TEC and its contribution to the GPS TEC, J. Atmos. Solar-Terr. Phys., 70, 1541-1548, (2008).

[31] E.V. Mishin, W.J. Burke and A.A. Viggiano, Stormtime subauroral density troughs: Ion-molecule kinetics effects, J. Geophys. Res., 109, A10301, (2004).

[32] B.T. Tsurutani, et al., Corotating solar wind streams and recurrent geomagnetic activity: A review, J. Geophys. Res., 111, A07S01, (2006).